%% file: main.tex
\documentclass[aip,cha,reprint,floatfix,amsmath,amssymb]{revtex4-2}
\usepackage[T1]{fontenc}
\usepackage{lmodern}
\usepackage{graphicx,booktabs,bm}
\usepackage{tikz}
\usetikzlibrary{arrows.meta,calc}
\usepackage[hidelinks]{hyperref}
\usepackage{amsthm}
\newtheorem{proposition}{Proposition}
\newcommand{\diag}{\operatorname{diag}}
\newcommand{\R}{\mathbb R}
\newcommand{\norm}[1]{\lVert#1\rVert}
\newcommand{\sinc}{\operatorname{sinc}}
\begin{document}
\title{Learning memory forces and bounding frequency errors in fractional Kuramoto grids}
\author{Hwanseo Lee}
\email{hwanbrother@kentech.ac.kr}
\affiliation{Department of Energy Engineering, Korea Institute of Energy Technology, Naju 58330, Republic of Korea}
\author{Hyunju Kim}
\email{hjkim@kentech.ac.kr}
\thanks{Corresponding author. Tel.: +82-61-320-9240.}
\affiliation{Department of Energy Engineering, Korea Institute of Energy Technology, Naju 58330, Republic of Korea}
\date{11 November 2026}
\begin{abstract}
Missing history creates two different demands on a forecaster for a fractional grid:
predicting frequency accurately and reconstructing memory well enough to bound
the prediction error. We connect these demands through the hereditary force
that remains after an integral observation correction. A compact decoder
retains the directions most relevant to that force, and an exact decomposition
separates omitted directions from coefficient error. A response-trained linear
predictor and neural refinement learn the coefficients for frequency prediction.
A separate force reconstruction supplies the reference for a nonlinear energy
bound for the specified finite relaxation model, accounting for numerical
propagation defects and the difference between the reference and point forecasts. For the outage transients studied here, refinement improves on the
linear predictor across the time steps and memory parameters tested. Replacing
the noisy origin states reveals why this gain can coexist with greater
memory reconstruction error: the learned correction partly compensates for observation
error. Separating prediction from reconstruction enables frequency bounds in
cases where the point predictor alone cannot support them. A critical radius
compares calibrated uncertainty with the angle margin available along the
reference trajectory, explaining both issuance and abstention. Under jointly
exchangeable calibration and test events, the calibrated guarantee bounds
the marginal probability of issuing an invalid envelope. Together, these
results connect uncertainty about history outside the observation window
to frequency forecasts and explicit conditions for bounding their errors.

\end{abstract}
\maketitle
\begin{center}
\footnotesize
\textcopyright\ 2026 Hwanseo Lee and Hyunju Kim.
This preprint is licensed under the Creative Commons Attribution--NonCommercial 4.0 International License
(\url{https://creativecommons.org/licenses/by-nc/4.0/}).
This license does not limit the separate distribution rights granted to arXiv.
\end{center}

\begin{quotation}
A fractional grid remembers motion that occurred before its current
observations. That hidden history matters through the force it exerts on the
future trajectory. Learning this force improves frequency prediction, but the
best predictor can compensate for noisy phase and frequency by altering its
memory estimate. Such compensation helps the forecast without reconstructing
the physical past more accurately. We use separate learning objectives for
these two tasks: one predicts frequency, and the other supplies a reference
for bounding its error. An energy argument connects the remaining memory
uncertainty to frequency, while an angle test determines whether the bound
can be issued throughout the forecast under the stated model and calibrated
uncertainty.

\end{quotation}

\section{Introduction}

Synchronization in a power network depends on its connectivity, power injections, and dynamical response. Kuramoto models provide a tractable description of this interaction, from collective phase coherence to the nonlinear response of an electrical network.\cite{acebron2005,rodrigues2016,filatrella2008,dorfler2012,dorfler2012grid,dorfler2013} Network heterogeneity and line disturbances also shape stability and frequency excursions.\cite{motter2013,menck2014,witthaut2022,schafer2018} A useful forecast must represent the states responsible for that response, including states that are absent from the instantaneous phase and frequency.

Fractional dynamics make this issue explicit. A power-law kernel weights the entire previous trajectory, and the start of an observation window is generally later than the start of the physical history. Physically coherent fractional initialization and observer-based reconstruction of the past have established precedents.\cite{sabatier2010,trigeassou2011} Fractional Kuramoto synchronization and its slow relaxation are also well established.\cite{ha2018} Recent direct-phase Caputo models further show how memory exponents change the admissible synchronized motion.\cite{das2026} Hybrid fractional networks have been studied with topology changes accompanied by deletion of earlier phase memory.\cite{ha2024} In a grid whose devices continue to operate through a line outage, changing the network torque and deleting the devices' memory are distinct operations. This distinction creates an initialization problem for a forecasting model.

Physics-informed machine learning offers ways to estimate hidden states and trajectories from data. Governing equations have been used in neural differential models, physics-informed neural networks, and fractional extensions.\cite{chen2018,rackauckas2020,raissi2019,karniadakis2021,pang2019} Operator learning extends this approach to mappings between input functions and trajectories,\cite{lu2021,li2021fno,kovachki2023,wang2021operator} including power-system component dynamics.\cite{karampinis2025} Graph models incorporate spatial interactions,\cite{sanchez2020,pfaff2021} and recent grid studies combine physics with probabilistic stability prediction and fractional graph propagation.\cite{pignn2026,akguller2026} Temporal fractional graph-feature dynamics also have direct precedents.\cite{kang2024} These developments motivate a graph-based inference model, but they leave the initial hereditary state to be specified or learned.

Several established approaches bear directly on the missing-history problem.
Interval observers bound unknown initialization in linear fractional
systems,\cite{frej2020} while learned nonlinear observer maps have robustness
analysis\cite{niazi2025} and computable approximation-error bounds.\cite{berincostain2026}
Fractional neural observers also address joint state and attack estimation.\cite{abdellah2026}
Goal-oriented inversion selects output-relevant directions for linear-Gaussian
inverse problems,\cite{spantini2017} and Bregman storage provides nonlinear
incremental energy estimates for microgrids.\cite{depersis2018}
Here the output is the inherited force remaining after an observation
correction and memory decay. Its low-rank approximation uses established
principles; the distinction is how separate prediction and reconstruction
objectives connect that force to a nonlinear trajectory bound with numerical
defect and angle constraints. Hidden-state learning and temporal pretraining offer other
ways to exploit dynamical observations.\cite{ozalp2023,ekambaram2026}
Recent work also places latent memory and exponential history compression
inside fractional neural representations,\cite{kumar2026} while trajectory-level
physics-informed learning regulates Kuramoto synchronization.\cite{luo2026}
These results establish that combining fractional dynamics with neural
learning is already an active approach; the question here is how an inferred
memory force can support a nonlinear frequency error bound.

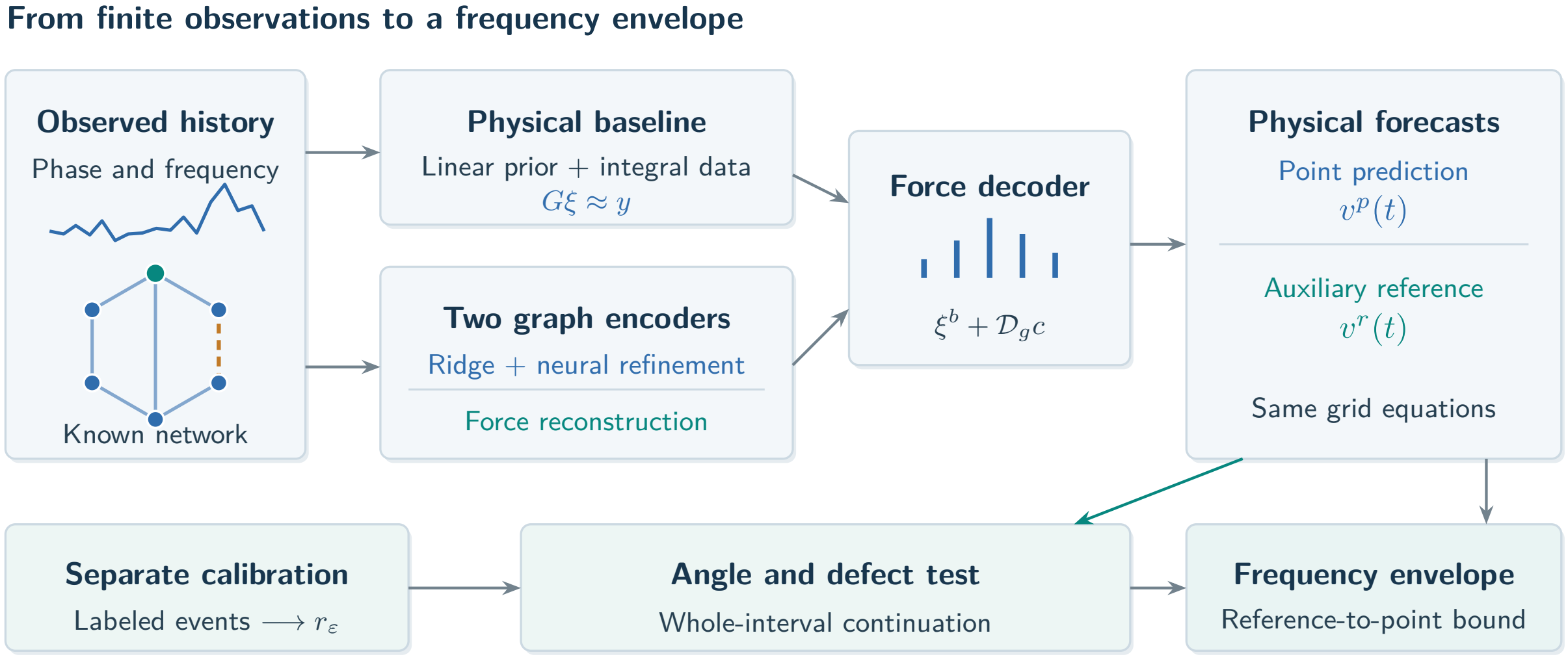
\begin{figure*}[!htbp]
\centering\resizebox{.94\textwidth}{!}{\input{figure/framework.tex}}
\caption{Physical and learned information meet in a compact force decoder.
The observed window supplies an integral baseline and inputs for learning
coefficients for prediction or force reconstruction. The resulting physical
forecasts serve distinct roles. Bound issuance depends on a radius calibrated
using the frozen reconstruction, an angle test, and the numerical defect.
The known difference between the reference and point forecasts transfers
the resulting envelope to the point forecast.}
\label{fig:framework}
\end{figure*}

The unresolved question is whether the memory estimate that best predicts
frequency is also the right reference for a physical error bound. We address
it in three connected steps. First, the observation equation determines which
memory corrections are compatible with a finite window; their future force
then defines a compact learning space. Second, a response-trained ridge
predictor and neural residual use that space to improve frequency forecasts.
Third, a separately trained force reconstruction supports a moving-reference
energy inequality, and calibration supplies the uncertainty radius it must
propagate. Figure~\ref{fig:framework} connects these steps. The experiments
test the resulting distinction directly: a forecast can improve even as its
memory reconstruction worsens, whereas an auxiliary reconstruction can enable
a bound that the point predictor cannot support. The critical-radius criterion
turns this distinction into an event-specific test for issuing the bound.

\section{Fractional grid dynamics and inherited history}

\subsection{Model and physical interpretation}

Consider an undirected lossless network without self-loops. Its arbitrarily oriented incidence matrix is $B\in\R^{N\times E}$, and $A=\diag(a_e)$ contains the positive strengths of active lines. In a fixed nominal synchronous frame, unwrapped phase $\theta_i$ and angular-frequency deviation $v_i$ satisfy
\begin{align}
 \dot\theta&=v,\label{eq:theta}\\
 M\dot v&=p+w-BA\sin(B^\top\theta)-Dv-Qm.
 \label{eq:grid}
\end{align}
Here $M,D,Q$ are diagonal, $M_i,D_i,q_i>0$, $p$ is a balanced scheduled injection, and $w$ is a specified disturbance. Time is nondimensional: with $\tau=t/T_0$ and power scale $P_0$, the dimensional coefficients become $M^{\rm dim}/(P_0T_0^2)$, $D^{\rm dim}/(P_0T_0)$, and $Q_i^{\rm dim}/(P_0T_0^{\alpha_i})$. Physical frequency deviation is $v_i/(2\pi T_0)$. Dots denote derivatives with respect to $\tau$, written as $t$ below.

Fractional damping is introduced through
\begin{equation}
 m_i(t)=b_i(t)+\int_0^t k_{\alpha_i}(t-s)v_i(s)\,ds,
 \quad k_\alpha(t)=\frac{t^{-\alpha}}{\Gamma(1-\alpha)},
 \label{eq:kernel}
\end{equation}
where $0<\alpha_i<1$ and $b_i$ is the contribution from history preceding the time origin. When that history is empty, $m_i={}^{C}_{0}D_t^{\alpha_i}\theta_i$, the Caputo derivative.\cite{diethelm2010} Ordinary frequency kinematics in Eq.~\eqref{eq:theta} remain intact: the fractional term acts as a hereditary damping force.

This is a reduced fixed-voltage swing model. The nodes describe dynamic machines or virtual oscillator units; a bus-voltage phasor requires a separate measurement relation. The inference study assumes observations of all node phases and frequencies and known constitutive parameters. These observations determine the integrated momentum relation used to initialize the hereditary state.

\subsection{Relaxation modes and a line event}

The kernel has the Laplace representation
\begin{align}
 k_\alpha(t)&=\int_0^\infty e^{-\lambda t}\mu_\alpha(\lambda)\,d\lambda,\nonumber\\
 \mu_\alpha(\lambda)&=\frac{\sin(\pi\alpha)}{\pi}\lambda^{\alpha-1}.
 \label{eq:measure}
\end{align}
Diffusive realizations of this form are established for fractional damping.\cite{deu2010} Their numerical use is connected to fractional predictor--corrector methods and fast history approximations.\cite{diethelm2002,diethelm2004,garrappa2018,jiang2017} With fixed rates $\lambda_r>0$ and positive quadrature weights $a_{ir}$, we use
\begin{equation}
 \dot z_{ir}=-\lambda_rz_{ir}+v_i,\qquad
 m_i^{(R)}=\sum_{r=1}^{R}a_{ir}z_{ir}.
 \label{eq:cells}
\end{equation}
Midpoint quadrature in $\ell=\log\lambda$ gives $a_{ir}=(\sin\pi\alpha_i/\pi)\lambda_r^{\alpha_i}\Delta\ell$. Write $z_r=(z_{1r},\ldots,z_{Nr})^\top$ and $A_r=\diag(a_{1r},\ldots,a_{Nr})$, so that $m^{(R)}=\sum_r A_rz_r$. The weights are determined by the fractional order and the quadrature, while $q_i$ controls amplitude. An independently adjustable weight vector would describe a more general kernel.

The initial condition $z_{ir}^0$ represents inherited history. In the continuum, a bounded prehistory on $[-H,0]$ gives $z_i^0(\lambda)=\int_{-H}^{0}e^{\lambda s}v_i(s)ds$. Admissible continuum fields require a defined output and finite storage and dissipation integrals. The finite representation in Eq.~\eqref{eq:cells} makes these quantities explicit. At a line event, $A$ changes and $(\theta,v,z)$ remain continuous when no device memory is reset.

The finite representation approximates the power-law response over a specified time range. Positive weights establish an energy property; accuracy also depends on the rate band and quadrature. Let $\epsilon_{\rm low}$ and $\epsilon_{\rm high}$ denote the $L^1(0,T)$ norms of the excluded low- and high-rate kernel contributions. They obey
\begin{align}
 \epsilon_{\rm low}&\leq \frac{\sin\pi\alpha}{\pi}\frac{T\lambda_{\min}^{\alpha}}{\alpha},\nonumber\\
 \epsilon_{\rm high}&\leq \frac{\sin\pi\alpha}{\pi}\frac{\lambda_{\max}^{\alpha-1}}{1-\alpha}.
 \label{eq:tails}
\end{align}
These bounds follow by integrating $(1-e^{-\lambda T})\mu_\alpha(\lambda)/\lambda$ over the discarded rates. Interior quadrature error must be added. The $L^1(0,T)$ kernel norm handles the integrable singularity at zero; uniform absolute error at zero is inappropriate. Neglect of fractional nonlocality can otherwise change the numerical problem.\cite{garrappa2019}

\subsection{Why the present phase and frequency are insufficient}

The missing history affects the acceleration immediately. This yields a local limitation for any predictor that uses only the present phase and frequency.

\begin{proposition}[Error from omitted memory]
\label{prop:memory}
Consider two finite realizations with identical parameters, the same locally Lipschitz future input on a fixed post-event topology, and the same initial $(\theta,v)$, but different memory states. Let $d=M^{-1}Q\sum_r A_r(z_r^{(1)}-z_r^{(2)})\ne0$. For small $h>0$, any common prediction $\widehat v(h)$ based only on the shared phase and frequency satisfies
\begin{equation}
 \max_{j=1,2}\norm{\widehat v(h)-v^{(j)}(h)}
 \geq \tfrac12h\norm{d}-Ch^2
 \label{eq:lower}
\end{equation}
for a finite local constant $C$.
\end{proposition}
\begin{proof}
The right-hand sides of Eq.~\eqref{eq:grid} differ initially only in the memory force, so $v^{(1)}(h)-v^{(2)}(h)=-hd+O(h^2)$. Smoothness holds for the finite realization on a fixed post-event topology. The triangle inequality implies that at least one error relative to a common prediction is at least half this separation, giving Eq.~\eqref{eq:lower}.
\end{proof}

The bound identifies the missing information. A finite observation window can constrain some combinations
of earlier motion; the following construction estimates the combinations
whose forces matter over the forecast horizon.
\section{Learning memory from integrated momentum}

\subsection{A boundary equation without acceleration estimates}

Let the observed interval begin at $s=0$ and end at $s=T_p$, the forecast origin. For each node,
\begin{align}
 z_{ir}(s)&=e^{-\lambda_rs}z_{ir}^{0}+z_{ir}^{f}(s),\nonumber\\
 z_{ir}^{f}(s)&=\int_0^s e^{-\lambda_r(s-\xi)}v_i(\xi)d\xi.
 \label{eq:split}
\end{align}
The observed velocity drives $z^f$, and $z^0$ contains the older history. Integrating Eq.~\eqref{eq:grid} at sample times $s_n$ gives a linear relation $H_i z_i^0=y_i$, with
\begin{align}
 (H_i)_{nr}&=a_{ir}\frac{1-e^{-\lambda_rs_n}}{\lambda_r},\label{eq:H}\\
 (y_i)_n&=-\frac{1}{q_i}\left[M_i\{v_i(s_n)-v_i(0)\}\right.\nonumber\\
 &\left.\quad+\int_0^{s_n}\{[BA\sin(B^\top\theta)]_i+D_iv_i-p_i-w_i\right.\nonumber\\
 &\left.\qquad+q_i\sum_r a_{ir}z_{ir}^{f}\}\,ds\right].
 \label{eq:y}
\end{align}
The relation uses velocity increments and integrals, avoiding numerical differentiation to estimate acceleration. In sampled data, the integrals and $z^f$ are evaluated from the available observations. A short window resolves combinations of modes more reliably than every individual mode, which motivates combining a prior with a physical correction.

\subsection{A reference correction and its linear prior}

Let $\xi_i=\diag(\sqrt{a_{ir}})z_i^0$ and $G_i=H_i\diag(a_{ir}^{-1/2})$. This weighting follows each mode's contribution to memory storage, apart from the node's common factor $q_i$. For a common fractional order, relaxation rates, and sampling schedule, $G$ is shared across nodes; write $a_r=a_{ir}$ in this setting. Although the realization contains many modes, they integrate the same velocity history and consequently vary together.

From training histories, pool the node vectors $\xi_s$ and form their mean $\mu$ and covariance
\begin{equation}
 C=\frac{1}{S-1}\sum_{s=1}^{S}(\xi_s-\mu)(\xi_s-\mu)^\top
   =U\diag(\nu_j)U^\top,
 \label{eq:covariance}
\end{equation}
with $\nu_1\geq\nu_2\geq\cdots\geq0$. When $\nu_1>0$, define
\begin{equation}
 L_r=U_r\diag\!\left(\sqrt{\nu_j/\nu_1}\right)_{j=1}^r.
 \label{eq:basis}
\end{equation}
The normalization separates the overall correction strength from the relative variation of the modes. The eigenvector truncation is principal component analysis (PCA),\cite{jolliffe2016} and covariance-based state estimation has a long history.\cite{kalman1960} Here the covariance describes inherited relaxation states. If its variance is zero, we use the unrestricted correction $L=I$ rather than normalize by zero.

A regularized linear prior uses relative phase, velocity, the network torque
before and after the scheduled line event, inertia, damping, memory
amplitude, and time. It receives local and weighted-neighbor window features;
normalization statistics and regression coefficients are obtained from training events. Denote its weighted
prediction at the window start by $\xi^\ell$.

\subsection{A regularized integral correction}

For $F=GL_r$ and $b=y-G\xi^p$, the correction is obtained from
\begin{equation}
 u^c=\arg\min_u\{\norm{Fu-b}^2+\lambda_{\rm reg}\norm{u}^2\},
 \qquad \xi^c=\xi^p+L_ru^c.
 \label{eq:objective}
\end{equation}
This is quadratic regularization in the learned correction coordinates.\cite{hoerl1970} The unrestricted alternative takes $L=I$. Validation selects either this alternative or a covariance-reduced correction for both the deployed observer and its calibrated linear comparator.

Strict convexity for $\lambda_{\rm reg}>0$ gives the familiar ridge solution
\begin{equation}
 \xi^c=\xi^p+L_r\mathcal K_r(y-G\xi^p),\qquad
 \mathcal K_r=(F^\top F+\lambda_{\rm reg} I)^{-1}F^\top.
 \label{eq:dual}
\end{equation}
The singular values of $\mathcal K_r$, $\sigma_j/(\sigma_j^2+\lambda_{\rm reg})$, are at most
$1/(2\sqrt{\lambda_{\rm reg}})$. Since both the normalized basis and $L=I$ have norm
at most one, this bounds sensitivity to additive error in $y$ with the prior
and correction operator held fixed. Perturbations of the observed window can
also change the prior; Eq.~\eqref{eq:decomposition} retains both contributions.
The correction acts only within the column space of the selected basis.

For $y=G\xi^\star+\eta$, the estimate obeys
\begin{equation}
 \xi^c-\xi^\star=(I-L_r\mathcal K_rG)(\xi^p-\xi^\star)+L_r\mathcal K_r\eta.
 \label{eq:decomposition}
\end{equation}
The identity separates prior error from observation and quadrature error without requiring them to be statistically independent. That distinction matters because the neural prior and the integral equation use the same observation window. Rank and $\lambda_{\rm reg}=\rho\norm{F}_2^2$ are selected using validation forecasts of the linear observer. The linear comparator receives both unrestricted and reduced corrections.

The gain is evaluated by a singular-value decomposition of $F$, avoiding the condition-number increase from forming normal equations. Its factorization costs $O(Kr^2+r^3)$ for $K$ observation intervals after $F$ is assembled. Gain application costs $O(NKR+NKr+NRr)$ across $N$ nodes. Its factorization is cached for the common kernel and sampling schedule. The unrestricted observer uses the same spectral evaluation with $K$ singular directions when $K<R$. Finally, $z_{ir}(T_p)=e^{-\lambda_rT_p}z_{ir}^c+z_{ir}^f(T_p)$ initializes the physical forecast.

\subsection{Shared physical propagation}

All predictors advance the same finite system. For
$V_0=\frac12v^\top Mv+\sum_ea_e(1-\cos\delta_e)-p^\top\theta
+\frac12\sum_{i,j}q_ia_{ij}z_{ij}^2$, direct differentiation gives
\begin{equation}
 \dot V_0=v^\top w-v^\top Dv-\sum_{i,j}q_ia_{ij}\lambda_jz_{ij}^2.
 \label{eq:energybalance}
\end{equation}
The memory force cancels the corresponding storage cross term. This
energy structure also motivates physics-informed learning and neural grid
control.\cite{greydanus2019,wang2026}
We use a discrete-gradient integrator,\cite{mclachlan1999} with frequency and memory evaluated at the midpoint and line torque given by
$\varphi_e=\sin[(\delta_e^{n+1}+\delta_e^n)/2]
\sinc[(\delta_e^{n+1}-\delta_e^n)/2]$.
Here $\sinc(x)=\sin(x)/x$ for $x\ne0$ and $\sinc(0)=1$. The identity
$\varphi_e(\delta_e^{n+1}-\delta_e^n)=\cos\delta_e^n-\cos\delta_e^{n+1}$
then gives the discrete version of Eq.~\eqref{eq:energybalance}.
Eliminating the mode midpoints leaves a nonlinear solve over the node
frequencies. A line event occurs between steps and changes network potential
by $\sum_e(a_e^+-a_e^-)(1-\cos\delta_e)$ while phase, frequency and memory
remain continuous. This balance constrains propagation; it does not establish
that an inferred initial memory is accurate.

\input{force_method.tex}

\section{Numerical results and interpretation}

\subsection{Experiments that separate prediction from reconstruction}

Adaptive integration generates trajectories of the specified model with known
parameters. Five units of prehistory lie outside the observed window. Training
and validation networks are cycles with one chord; removing a cycle line
at the forecast origin keeps them connected. Parameter variation and Gaussian
initial perturbations produce different transients (Table~\ref{tab:design}).

\input{design_table.tex}
\input{reproducibility.tex}

Weights, amplitudes and reference choices were frozen before the final
calibration and test groups were generated. Events used during development for
method selection are excluded from these groups, which include larger cycles
and meshed networks. Each group has
separate calibration events from the same distribution. The noise level
is known; Gaussian perturbations affect observed phase and frequency. We call standard deviations $0.002$ and $0.03$ the primary- and high-noise conditions, respectively.
Calibration selects no model settings.

All comparisons share physical propagation and observed phase and frequency
at the forecast origin. The force and principal-direction comparison isolates
the decoder subspace. Direct-prior and state-residual comparisons instead
compare complete inference procedures, with different training objectives,
noise augmentation and correction choices.
The reference-memory comparator uses the true memory at the forecast origin
with the same noisy electrical states. Because an inferred memory can
compensate for errors in those states, the reference-memory error is not
a lower bound on forecast error.

Frequency root-mean-square error (RMSE) is computed over nodes and future samples
within each event, excluding the supplied origin. Neural comparisons average
five training repetitions within each event. Paired bootstrap intervals
resample whole events and condition on the trained models. Bound experiments deploy one fixed training replicate. Their score coverage is marginal within each calibrated event group. Five disjoint calibration sets
are compared on the same test events; they are not five independent test
ensembles. Wilson intervals quantify finite-test binomial uncertainty
conditional on each calibration set. Observed tube failures are
checked at stored reference times; the analytic issuance test itself covers
whole interpolation intervals.

Trajectory generation and prediction use the same finite relaxation model.
The sensitivity study distinguishes variation within a realization from
convergence toward the power-law kernel.

\subsection{Force geometry reduces the learning dimension}

\begin{figure*}[!htbp]
\centering\includegraphics[width=.78\textwidth]{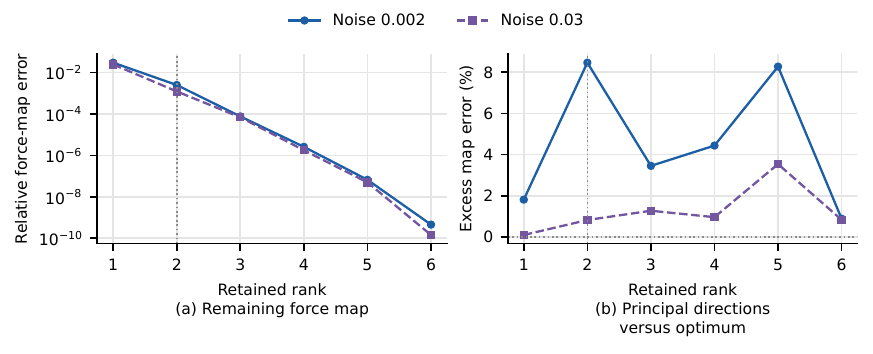}
\caption{Force-map compression at two observation-noise levels. The left
figure gives the relative spectral norm of the omitted force map under the
optimal force directions. The right figure shows the principal-direction comparator's
relative excess error over that optimum, $100(\epsilon_{\rm principal}/\epsilon_{\rm force}-1)$.
Both use the same training-error second moment, physical correction and rank.
The vertical reference marks the retained rank for these noise levels.}
\label{fig:spectrum}
\end{figure*}

The training force spectrum retains one coordinate for clean observations
and two for the noisy conditions. The correction and decay over the observation window have already
suppressed directions that are resolved by observations or decay rapidly. Ranking the remaining
directions by future force concentrates learning on the uncertainty that can
still affect the forecast. Figure~\ref{fig:spectrum} compares
the corresponding force-map errors with principal directions at equal rank.
The singular-value result explains this difference within the specified map;
the frequency comparisons below test its consequence for nonlinear prediction.

\subsection{A strong linear decoder and its nonlinear refinement}

\input{accuracy_table.tex}

Refinement reduces mean error relative to the response-trained ridge predictor by $3.6$--$6.2\%$ at primary noise and $7.0$--$8.2\%$ at the high-noise level. All eight individual paired confidence intervals against this ridge control exclude zero in favor of refinement. Reductions versus direct graph priors at primary noise are $30.0$--$38.4\%$; these compare complete procedures rather than isolating architecture. The principal-direction refinement gives a more discriminating comparison. At primary noise, the force refinement improves the six-node and ten-cycle forecasts, while the two meshed-group intervals span zero and their principal-direction point estimates are slightly lower. At the largest noise level, force-coordinate reductions are $3.3$--$3.9\%$ and all four paired intervals favor force coordinates. Both decoders capture the dominant history contribution at lower noise. The high-noise result indicates that choosing the remaining directions for post-correction force becomes more useful as initial-state errors grow; the exact map optimum and the empirical frequency ordering remain distinct results.

At the two tabulated noise levels, the stand-alone neural coefficient predictor is less accurate than the
response-trained ridge predictor, while neural refinement improves its predictions.
Keeping the linear solution as an anchor lets the network focus on the
remaining frequency error.
The state-residual graph prior is a separate control that corrects the full
memory state. Its validation gain at primary noise falls below its
predeclared numerical-resolution threshold, so its amplitude is zero and
its table entry equals the corrected linear prior. The force refinement
uses the response objective and its own validation amplitude; the time-step refinement comparison tests whether its smaller gains exceed numerical propagation effects.

\input{refinement_results.tex}

\begin{figure*}[!htbp]
\centering\includegraphics[width=.74\textwidth]{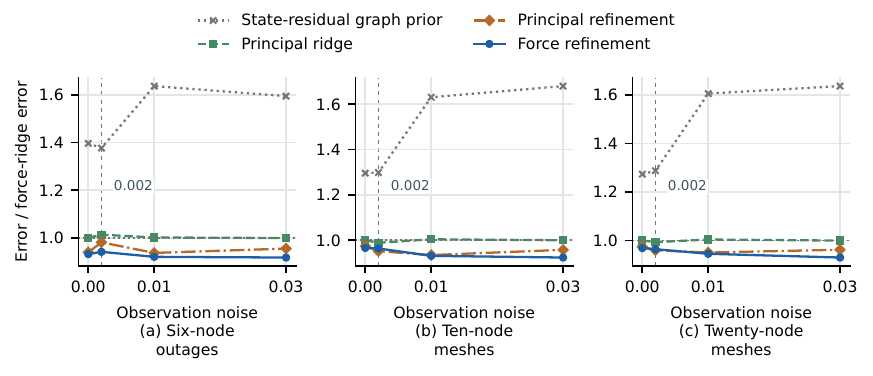}
\caption{Mean frequency error relative to the force-coordinate response-trained
ridge decoder. The horizontal reference line denotes equal error; the vertical dashed marker
identifies the primary-noise setting. The plots use new
six-node, ten-node meshed, and twenty-node meshed events. Curves distinguish
the state-residual graph prior, principal-direction controls, and the proposed
force-coordinate refinement.}
\label{fig:prediction}
\end{figure*}

\subsection{Prediction and reconstruction support different bounds}

The gain in prediction raises a physical question: has the model inferred
the hidden memory more accurately, or has it used the memory estimate to
compensate for another error? The decomposition and origin-state intervention answer this question
without changing the memories inferred from the observation windows.

\input{mechanism_results.tex}
\begin{figure*}[!tb]
\centering\includegraphics[width=.94\textwidth]{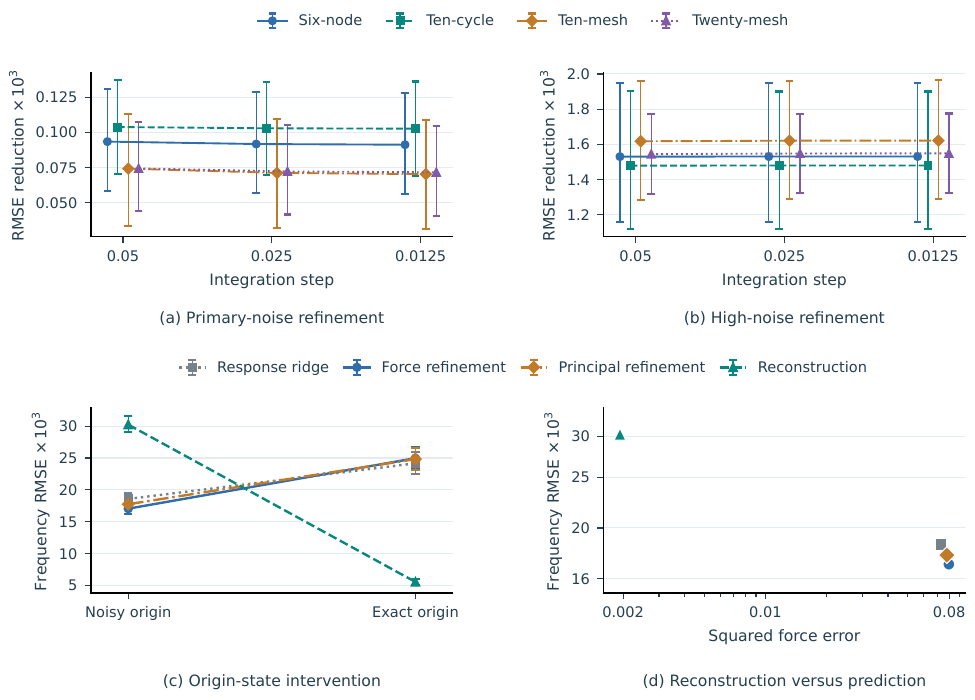}
\caption{Separating numerical convergence from error compensation.
(a,b) Mean reduction in frequency root-mean-square error (RMSE) relative to
the response ridge at three integration step sizes; vertical intervals are paired
$95\%$ event-bootstrap intervals, conditional on five trained refinements.
Horizontal offsets separate groups with the same step. (c) Six-node high-noise
forecasts with the same inferred memories, before and after replacing only
the phase and frequency at the forecast origin with their exact values. Lines join means, with
paired-event bootstrap intervals for each mean. (d) Mean squared inherited-force
error against mean frequency RMSE on those noisy-origin events. Each point
represents one learning objective; axes are logarithmic. The force decomposition
in Table~\ref{tab:mechanism} explains why refinement can improve prediction
while increasing memory-force error.}

\label{fig:diagnostics}
\end{figure*}

The intervention identifies compensation of origin-state errors as a
mechanism behind the forecast gain. A bound based on the separate magnitudes
of those errors cannot rely on their cancellation. We consequently retain
the frequency-trained point prediction and propagate uncertainty around a
force-trained reference, adding their known discrepancy through
Eq.~\eqref{eq:transport}. The next comparison tests whether this separation
improves the availability of bounds.

\input{certificate_table.tex}

Table~\ref{tab:certificate} quantifies envelope width and calibration variability.
Both auxiliary references issue bounds for nearly all
primary-noise events. At high noise, separating the reference from the point
predictor enables issuance in three groups; the point reference issues none.
The neural reference adds only a few events relative to the linear reference,
and this difference is smaller than the variation across calibration sets.
Reference separation accounts for most of the gain in issuance; the choice between linear and neural reconstruction has a smaller effect here. No issued physical bound fails at the stored times.

The trajectory-residual comparator uses the same point predictor and label
budget, calibrating $\max_j\|v(t_j)-v^p(t_j)\|$ directly. It supplies a
constant envelope at the stored times with marginal coverage at least
$1-\varepsilon$. Its narrower envelopes are useful when the target is this
sampled trajectory. The wider physics-based envelope reflects continuous-time
propagation, a force-based score, and a sufficient angle condition. Neither
the smaller direct envelope nor its empirical coverage establishes a
continuous-time guarantee between the samples.

\input{calibration_interpretation.tex}

\begin{figure*}[!htbp]
\centering\includegraphics[width=.78\textwidth]{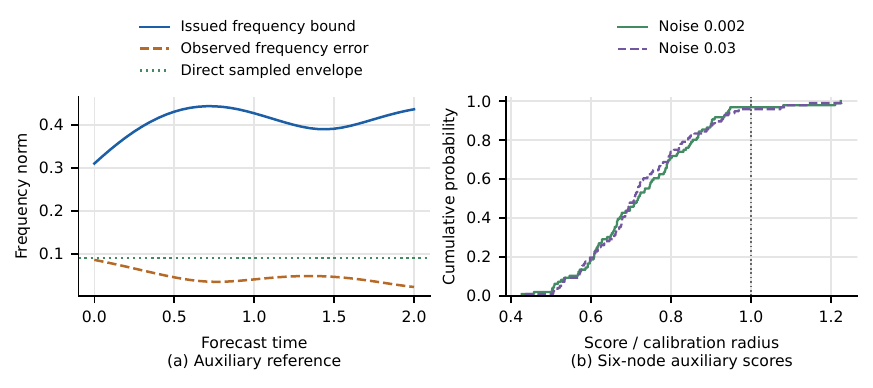}
\caption{Nonlinear frequency bounds and calibration. The left figure shows
an auxiliary-reference bound for the first six-node high-noise event where
the point reference abstains. Selection uses the issuance test, not future
error; the dotted line is the direct sampled-time envelope. The right figure
shows test scores normalized by their respective calibration radii. The
vertical line marks unit score, not conditional coverage. Curves show stored
times; the issuance test covers whole intervals.}
\label{fig:tube}
\end{figure*}

\begin{figure*}[!tb]
\centering\includegraphics[width=.90\textwidth]{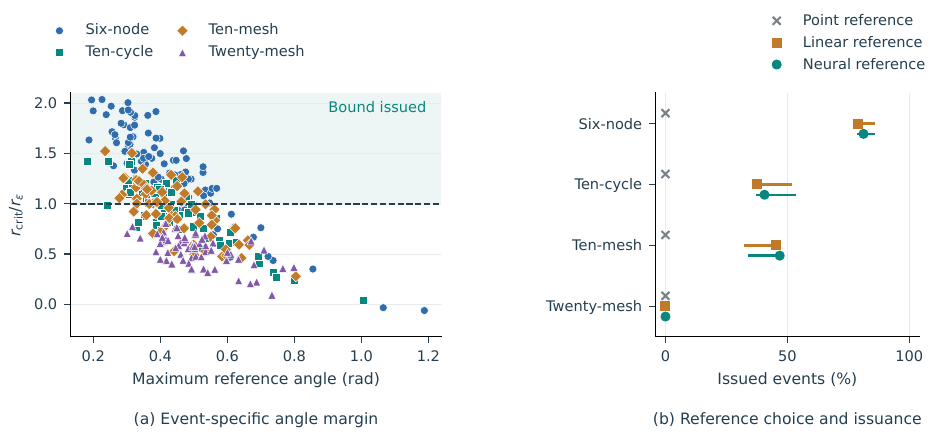}
\caption{When a reconstruction reference permits a frequency bound.
(a) Every high-noise test event under the first calibration set, using the
neural reconstruction reference. The horizontal line marks the strict issuance
threshold $r_{\rm crit}/r_\varepsilon=1$; shaded values exceed it. The horizontal
axis is the largest upper bound on the absolute reference angle over active edges and time.
Each group uses its own calibration radius. (b) Percentage of events issued
a bound by each reference choice. Markers denote the first calibration set;
horizontal segments span five disjoint calibration sets evaluated on the
same test events. These spans describe calibration variability, not confidence
intervals. Linear and neural references share the same point predictor.}
\label{fig:issuance}
\end{figure*}

A successful bound depends on the position of the calibrated radius relative
to the event's critical radius. Figure~\ref{fig:issuance} displays this geometry
for all high-noise test events. At comparable reference angles, events can
have different margins because growth, effective resistance and numerical
defect also enter Eq.~\eqref{eq:critical}. The twenty-node group lies below
the issuance threshold throughout. This connects abstention to the sufficient
error test without interpreting it as a loss of physical synchrony.

\input{sensitivity_results.tex}
\subsection{Computation and numerical consistency}

The critical-radius calculation reuses each reference path across calibration radii.
The force geometry and observer gain are reused under common kernel and
sampling conditions. The point predictor estimates few decoder coordinates;
when validation sets its nonlinear residual amplitude to zero, the neural evaluation is
skipped. Table~\ref{tab:cost} separates initialization from the full point forecast.
Issuing a bound additionally requires a refined reference and whole-interval
defect and angle estimates; caching avoids repeating those calculations.

\input{cost_table.tex}
\input{cached_timing.tex}

The reference-memory frequency error decreases from
$1.59\times10^{-4}$ to $9.95\times10^{-6}$ as the step decreases from
$0.05$ to $0.0125$, consistent with second-order convergence.
Tightening adaptive-reference tolerances on development events changes
frequency by at most $1.11\times10^{-7}$, well below this propagation error.
Across the final point comparisons, the largest energy defect is
$2.64\times10^{-15}$ and the largest equation residual is $2.49\times10^{-15}$.
Separate checks compare the force Gramian with direct integration,
the decoder update with a fresh physical correction, and the exact affine-drive
mode response with a matrix exponential. They also test the Pythagorean
identity, moving-reference remainder, and whole-interval defect and angle
enclosures. Halving the response perturbation tests the local training
derivative. These checks distinguish physical propagation, memory
reconstruction and numerical error.

\section{Discussion and Conclusion}

The results identify inherited force as the connection between an unobserved
past and a useful grid forecast. Applying the observation correction before
constructing the force decoder concentrates learning on the uncertainty that
survives into the forecast. Neural refinement improves a strong linear
predictor in the tested regimes, and the gain persists under time-step refinement.
The origin-state intervention gives this result a sharper interpretation:
the learned memory correction can compensate for electrical observation error,
so prediction accuracy and physical-memory accuracy need not improve together.

This separation of objectives is also the reason for using an auxiliary
reconstruction in the error bound. Its calibrated force uncertainty propagates
through the nonlinear energy inequality, while the reference-to-prediction
discrepancy keeps the envelope attached to the chosen point forecast. The
critical radius expresses the available angle margin as a scalar test. It
explains why the same method issues bounds for some outages and abstains for
others, and permits repeated calibration without recomputing the reference
path. Direct residual calibration gives tighter sampled-time envelopes;
the energy construction instead supplies a sufficient whole-interval statement
for the declared finite realization. Across issued physical bounds, the
median maximum-envelope/error ratios range from $5.8$ to $14.6$, compared
with $1.2$--$1.6$ for direct sampled-time calibration; the high-noise
twenty-node group receives no physical bounds. This conservatism reflects
discarded dissipation, a horizon-wide force norm and a sufficient angle
condition. The separate reference enables issuance without ensuring a
narrow envelope, and the linear reference accounts for most of that benefit.

The experiments use known parameters, observations at all nodes, a common
fractional order and labeled simulated histories. Kernel refinement shows
that realization error can exceed the differences between estimators, making
physical-model resolution as important as learning accuracy. Unknown order
also requires separating its effect from damping, amplitude and initial
memory; information-rank analysis and profile likelihood offer tools for
this extension.\cite{rothenberg1971,raue2009} Optimization and causal loss
construction remain relevant when training in more difficult
regimes.\cite{wang2021gradient,krishnapriyan2021,wang2022causal}
Field frequency measurements, multiscale data and varied dispatch scenarios
provide routes to testing the observation and parameter models beyond these
simulations.\cite{kruse2023,zheng2022,gillioz2025} Extending the probability
statement to such settings requires calibration events representative of the
new operating conditions.

\section*{Acknowledgments}
This work was supported by the National Research Foundation of Korea (NRF) grant funded by the Korean government, Ministry of Science and ICT (MSIT), under Grant No. RS-2026-25589376.

\section*{Author declarations}
\subsection*{Conflict of interest}
The authors have no conflicts to disclose.
\subsection*{Author contributions}
Hwanseo Lee: Data curation (equal); Investigation (equal); Methodology (equal);
Validation (equal); Visualization (lead); Writing---review \& editing (equal).

Hyunju Kim: Conceptualization (lead); Data curation (equal); Investigation (equal);
Methodology (equal); Funding acquisition (lead); Supervision (lead);
Validation (equal); Visualization (equal); Writing---original draft (equal);
Writing---review \& editing (equal).

\section*{Data availability}
The complete reproducibility package for this manuscript version, including
the data, simulation code, trained model weights, selection-provenance inputs
and replay and retraining instructions, is available from the corresponding
author upon reasonable request.
\bibliographystyle{aipnum4-2}
\bibliography{references}
\end{document}

%% file: figure/framework.tex
% Editable TikZ source for Figure 1. All coordinates are in centimetres.
\begingroup
\definecolor{gridnavy}{HTML}{17354D}
\definecolor{gridblue}{HTML}{306DAD}
\definecolor{gridteal}{HTML}{07877F}
\definecolor{gridamber}{HTML}{C17A27}
\definecolor{gridink}{HTML}{263F50}
\definecolor{gridline}{HTML}{CCD9E2}
\definecolor{gridpale}{HTML}{F3F7FA}
\definecolor{gridmint}{HTML}{EDF6F4}
\begin{tikzpicture}[x=1cm,y=1cm,
  font=\sffamily\fontsize{8.5}{10}\selectfont,
  heading/.style={font=\sffamily\bfseries\fontsize{9.5}{11}\selectfont,text=gridnavy},
  card/.style={rounded corners=3pt,draw=gridline,line width=.65pt,fill=gridpale},
  lower/.style={card,fill=gridmint},
  flow/.style={-{Stealth[length=2mm,width=1.5mm]},draw=gridink!65,line width=.85pt},
  ref flow/.style={flow,draw=gridteal},
  every node/.style={inner sep=0pt,text=gridink,align=center}]
\path[use as bounding box] (-.05,-.05) rectangle (16.65,6.95);
\node[anchor=west,heading,font=\sffamily\bfseries\fontsize{10}{11}\selectfont]
  at (0,6.73) {From finite observations to a frequency envelope};

% Solid, pale shadows keep the artwork compatible with vector EPS export.
\foreach \x/\y/\w/\h in {0/2.05/3.2/4.15,4/4.55/4.4/1.65,
  4/2.05/4.4/2.05,9/3.05/3/2.5,12.6/2.05/4/4.15,
  0/0/4.3/1.35,5.5/0/6.5/1.35,12.6/0/4/1.35}{
  \path[rounded corners=3pt,fill=gridline!50,draw=none]
    (\x+.035,\y-.045) rectangle (\x+\w+.035,\y+\h-.045);
}
\path[card] (0,2.05) rectangle (3.2,6.2);
\path[card] (4,4.55) rectangle (8.4,6.2);
\path[card] (4,2.05) rectangle (8.4,4.1);
\path[card] (9,3.05) rectangle (12,5.55);
\path[card] (12.6,2.05) rectangle (16.6,6.2);
\path[lower] (0,0) rectangle (4.3,1.35);
\path[lower] (5.5,0) rectangle (12,1.35);
\path[lower,fill=gridteal!9] (12.6,0) rectangle (16.6,1.35);

\node[heading] at (1.6,5.62) {Observed history};
\node at (1.6,5.12) {Phase and frequency};
% Schematic trace, not an additional experimental result.
\draw[gridblue,line width=1pt,line join=round] plot coordinates {
  (.47,4.48) (.62,4.45) (.75,4.54) (.9,4.44) (1.03,4.59)
  (1.17,4.38) (1.31,4.45) (1.46,4.46) (1.61,4.51)
  (1.76,4.49) (1.9,4.63) (2.04,4.46) (2.19,4.79)
  (2.34,4.98) (2.48,4.7) (2.63,4.75) (2.76,4.48)};
\begin{scope}[shift={(1.6,3.25)},scale=.78]
  \foreach \n/\a in {1/90,2/150,3/210,4/270,5/330,6/30}
    \coordinate (g\n) at (\a:1);
  \foreach \a/\b in {1/2,2/3,3/4,4/5,6/1,1/4}
    \draw[gridblue!60,line width=1pt] (g\a)--(g\b);
  \draw[gridamber,line width=1.25pt,dash pattern=on 3pt off 2pt] (g5)--(g6);
  \foreach \n in {2,...,6} \filldraw[fill=gridblue,draw=white,line width=.7pt] (g\n) circle (3.3pt);
  \filldraw[fill=gridteal,draw=white,line width=.7pt] (g1) circle (3.8pt);
\end{scope}
\node at (1.6,2.32) {Known network};

\node[heading] at (6.2,5.62) {Physical baseline};
\node at (6.2,5.15) {Linear prior + integral data};
\node[text=gridblue] at (6.2,4.78) {$G\xi\approx y$};

\node[heading] at (6.2,3.53) {Two graph encoders};
\node[text=gridblue] at (6.2,3.03) {Ridge + neural refinement};
\draw[gridline,line width=.6pt] (4.3,2.79)--(8.1,2.79);
\node[text=gridteal] at (6.2,2.46) {Force reconstruction};

\node[heading] at (10.5,4.97) {Force decoder};
\foreach \x/\h in {9.8/.20,10.15/.40,10.5/.64,10.85/.47,11.2/.27}
  \draw[gridblue,line width=1.7pt] (\x,3.98)--(\x,3.98+\h);
\node at (10.5,3.48) {$\xi^b+\mathcal D_g c$};

\node[heading] at (14.6,5.62) {Physical forecasts};
\node[text=gridblue] at (14.6,5.1) {Point prediction};
\node[text=gridblue,font=\fontsize{10}{11}\selectfont] at (14.6,4.69) {$v^p(t)$};
\draw[gridline,line width=.6pt] (12.95,4.34)--(16.25,4.34);
\node[text=gridteal] at (14.6,3.85) {Auxiliary reference};
\node[text=gridteal,font=\fontsize{10}{11}\selectfont] at (14.6,3.43) {$v^r(t)$};
\node at (14.6,2.57) {Same grid equations};

\node[heading] at (2.15,.79) {Separate calibration};
\node at (2.15,.31) {Labeled events $\longrightarrow r_\varepsilon$};
\node[heading] at (8.75,.79) {Angle and defect test};
\node at (8.75,.31) {Whole-interval continuation};
\node[heading] at (14.6,.79) {Frequency envelope};
\node at (14.6,.31) {Reference-to-point bound};

\draw[flow] (3.2,5.32)--(4,5.32);
\draw[flow] (3.2,3.03)--(4,3.03);
\draw[flow] (8.4,5.08)--(9,4.78);
\draw[flow] (8.4,3.05)--(9,3.65);
\draw[flow] (12,4.34)--(12.6,4.34);
\draw[flow] (4.3,.67)--(5.5,.67);
\draw[flow] (12,.67)--(12.6,.67);
\draw[ref flow] (13.2,2.05)--(11.4,1.35);
\draw[flow] (15.8,2.05)--(15.8,1.35);
\end{tikzpicture}
\endgroup

%% file: force_method.tex
\section{Learning memory that matters to the forecast}

The integral equation constrains an initial memory vector, whereas a forecast
responds to its force. These are different approximation objectives: an error
in a rapidly decaying mode can be large in state coordinates and negligible in its effect
on the forecast. We use this distinction to choose the neural decoder.

\subsection{The force geometry after correction}

Write $E_p=\diag(e^{-\lambda_rT_p})$, $J=I-L\mathcal K G$, where
$L$ and $\mathcal K$ denote the selected correction basis and gain, and
$\xi^b=E_p\xi^c+\xi^f$ for the weighted memory at the forecast origin
obtained from the linear prior. The superscript $b$ denotes this baseline,
and $\xi^f=\diag(\sqrt a)z^f(T_p)$. An increment $h$ in the prior at the
start of the observation window changes this origin state by $E_pJh$.
Let $C_\ell$ be the training second moment of the linear-prior error,
\begin{equation}
 C_\ell=\frac1S\sum_s(\xi_s-\xi_s^\ell)(\xi_s-\xi_s^\ell)^\top.
 \label{eq:errorsecond}
\end{equation}
Its uncentered definition includes the mean error.

For a forecast duration $T_f$, the exact Gramian of inherited memory output is
\begin{equation}
 W_{jk}=\sqrt{a_ja_k}\,
 \frac{1-e^{-(\lambda_j+\lambda_k)T_f}}{\lambda_j+\lambda_k}.
 \label{eq:forcegramian}
\end{equation}
For any origin-memory error $e$, $e^\top We$ equals
$\int_0^{T_f}|\sum_j\sqrt{a_j}e^{-\lambda_jt}e_j|^2dt$.
The node amplitude $q_i$ scales both this force and its norm without
changing the node's decoder directions.
We factor the correction-dependent operator
\begin{equation}
 S_F=W^{1/2}E_pJC_\ell^{1/2}
       =\mathsf U\Sigma\mathsf V^\top
 \label{eq:goaloperator}
\end{equation}
and define a prior basis and its origin decoder by
\begin{equation}
 \mathcal B_g=C_\ell^{1/2}\mathsf V_g,\qquad
 \mathcal D_g=E_pJ\mathcal B_g.
 \label{eq:forcedecoder}
\end{equation}
Only nonzero singular directions are retained. If $S_F=0$, the decoder has
rank zero and adds no neural correction. Singular subspaces with repeated
values admit equivalent rotations within a fully retained subspace. If the
cutoff splits a tied block, the optimal subspace need not be unique; we store
the chosen basis and use it consistently for training and inference.
The implementation selects the smallest $g$ satisfying
$\sum_{j>g}\sigma_j^2/\sum_j\sigma_j^2\leq10^{-4}$, restricting the retained
directions to $\sigma_j>10^{-8}\sigma_1$. The zero-map case is handled
separately. This energy criterion differs from the spectral-norm error
displayed in Fig.~\ref{fig:spectrum}. Changing the kernel, observation window
or forecast horizon requires rebuilding this operator and recalibrating
the score, as done in the sensitivity study.

This construction specializes established low-rank approximation and
goal-oriented inverse methods.\cite{eckart1936,spantini2017} Its role here is
to place learning after both observation-window inference and physical
memory decay, instead of compressing the uncorrected state alone.

\begin{proposition}[Force compression and the coefficient error]
For $e=E_pJC_\ell^{1/2}u$, the best rank-$g$ approximation of the map from
$u$ to inherited force in the $L^2(0,T_f)$ norm has worst-case unit-ball
error $\sigma_{g+1}(S_F)$, with $\sigma_{R+1}=0$. The decoder in Eq.~\eqref{eq:forcedecoder}
attains it. For an arbitrary actual origin error $e=\xi^{\rm true}-\xi^b$ (truth minus baseline), define
\begin{equation}
 c^*=\Sigma_g^{-1}\mathsf U_g^\top W^{1/2}e.
 \label{eq:forcecoeff}
\end{equation}
Every coefficient prediction $\widehat c$ obeys
\begin{align}
 \|W^{1/2}(e-\mathcal D_g\widehat c)\|^2
 &=\|(I-\mathsf U_g\mathsf U_g^\top)W^{1/2}e\|^2\nonumber\\
 &\quad+\|\Sigma_g(c^*-\widehat c)\|^2.
 \label{eq:forcedecomposition}
\end{align}
\end{proposition}
\begin{proof}
The force map is isometric to $S_F$ through its Gramian. The truncated singular-value
decomposition gives the first statement. Since
$W^{1/2}\mathcal D_g=\mathsf U_g\Sigma_g$, the two terms in the second
statement are orthogonal projections, giving the identity.
\end{proof}

The first term in Eq.~\eqref{eq:forcedecomposition} is an omitted-force
floor: increasing encoder capacity cannot remove it within the same decoder.
The identity uses the actual error, including observation noise and window
quadrature, without an independence assumption. Singular-value optimality
concerns the declared force map and uncertainty shape; it is not a theorem
of optimal nonlinear frequency prediction.

\subsection{Separate objectives for prediction and reconstruction}

A shared graph layer aggregates local and weighted-neighbor window features.
A causal temporal attention encoder\cite{vaswani2017} outputs the retained
coefficients, with the known observation-noise level as an additional input.
Feature means and scales are calculated from training events. A coefficient update can be applied
either before or after the physical correction:
\begin{equation}
 \xi^{p}=\xi^\ell+\mathcal B_g\widehat c
 \quad\Longleftrightarrow\quad
 \xi^{\rm origin}=\xi^b+\mathcal D_g\widehat c.
 \label{eq:commute}
\end{equation}
Thus the compact decoder uses the same regularized physical correction,
with $G$, $y$, and the penalty unchanged. This does not impose exact agreement
with noisy integrated data.

We distinguish a point predictor from a memory reconstruction used to build
an error bound. For each noise condition, let $e_{si}$ be the weighted
origin-memory error of the baseline at node $i$ in training event $s$.
The reconstruction objective is
\begin{align}
 \mathcal E_m&=\max\left\{\frac{1}{n_{\rm tr}N}
       \sum_{s,i} e_{si}^{\top}We_{si},10^{-20}\right\},\nonumber\\
 \mathcal L_m&=\frac{1}{N\mathcal E_m}\mathbb E_s
       \sum_i q_i^2\|W^{1/2}\mathcal D_g(c^*_{si}-\widehat c_{si})\|^2.
 \label{eq:reconstructionloss}
\end{align}
Here $n_{\rm tr}$ is the number of training events. The normalization uses
unweighted baseline force error; the floor prevents division by zero.
Training averages this loss equally across noise conditions, and validation
uses the same training normalizations. For point prediction, a local response
of the physical integrator supplies the differentiable objective
\begin{equation}
 \mathcal L_v=\frac1{Nn_f\mathcal E_v}
 \mathbb E\|\widehat{\mathbf v}^{\rm cent}
                  +\mathcal R\widetilde c-\mathbf v^{\rm ref}\|^2,
 \label{eq:responseloss}
\end{equation}
where $\widetilde c$ is standardized, $\widehat{\mathbf v}^{\rm cent}$ is
the forecast at the training coefficient mean, and $\mathcal R$ is its
coefficient response. Centered perturbations evaluate that response.
$n_f$ is the number of forecast samples and the expectation is over events.
For each noise condition, $\mathcal E_v$ is the corresponding baseline
frequency mean-square error: training uses the training baseline, while
validation checkpoint selection uses the validation baseline. Feature and
coefficient means and scales still come from training events.
This is a local training surrogate; checkpoint selection uses the validation surrogate, while final amplitude
selection and all test forecast errors use the nonlinear forecast. Reference frequency trajectories are used in training and validation; future test trajectories are used only for final scoring.

For point prediction, a ridge model first estimates coefficients using the
same response objective and local and neighbor features. The neural output
refines that linear prediction through a residual. Refinement reuses the local response at the original
training center; it does not recompute a response at each ridge prediction.
Its initial residual is zero. Validation can disable
it, in which case inference uses the ridge model alone.

The neural objectives use the same encoder size and adaptive moment estimation
(Adam) optimization schedule.\cite{kingma2014}
Checkpoint selection uses the corresponding validation objective. A final
amplitude is selected by nonlinear validation error for prediction and by
the mean validation-event norm $\|Qb_\beta\|_{L^2}$ for reconstruction,
where $b_\beta$ is the inherited error after scaling the predicted coefficient
by $\beta$. Table~\ref{tab:mechanism} instead summarizes the mean squared norm.
Calibration events select neither
weights nor amplitudes. The experiments retain the ridge-only predictor as
a control. For comparison, let $P_g$ contain the leading eigenvectors of the
same $C_\ell$. The basis $C_\ell^{1/2}P_g$ gives the principal-direction
control: it applies principal component analysis without centering the prior
errors. It uses the same correction, decoder rank, ridge objective and
neural training procedure. Thus the prior-error second moment is unchanged;
the subspace criterion alone differs.

The linear auxiliary reference uses the corrected linear prior of
Sec.~III, propagated to the forecast origin as $\xi^b$; it adds no coefficient
correction. Its ridge regression predicts standardized window-start memory
from the local and neighbor features by squared-error minimization with a
quadratic coefficient penalty. It is trained at the primary noise level;
validation forecasts at each noise level select its penalty and integral
correction. Thus it differs from the response ridge, which learns decoder
coefficients using $\mathcal L_v$. The linear reference has no extra amplitude
parameter. Both reference choices and the point predictor are frozen before
calibration; each reference has its own score radius within each network
and noise group.

Different objectives are needed because a memory correction can be
correlated with errors in the initial phase and velocity. A coefficient that improves frequency
need not be the most accurate reconstruction of physical memory. The bound
below can use the reconstruction as an auxiliary reference and then account
explicitly for its difference from the point forecast. Direct-prior controls
also include a gated recurrent unit,\cite{cho2014} attention, and selective
exponential recurrence. Long short-term memory,\cite{hochreiter1997}
state-space layers,\cite{gu2022} and selective state-space architectures\cite{gu2023,dao2024}
are established sequence-model precedents; no new attention or recurrence
architecture is claimed.

\section{From memory-force error to nonlinear frequency bounds}

The compression identity explains what learning can change. To determine
what the remaining force error means for a grid forecast, we compare the
true motion with a moving reference, rather than linearizing about an
equilibrium. Bregman storage is established in microgrid
analysis.\cite{depersis2018} Here it gives a finite-horizon relation between
inherited memory and frequency, including a numerical reference defect.

\subsection{A moving-reference energy inequality}

After the line event, fix the connected active topology and positive device
coefficients, and take $w=0$ over the forecast as in the experiments. Here
$B$ and $A$ refer to active lines. Let $(\theta^r,v^r,z^r)$ be an absolutely continuous reference
with $\dot\theta^r=v^r$ and exact relaxation equations. Its momentum defect is
\begin{equation}
 d=M\dot v^r-p+BA\sin(B^\top\theta^r)+Dv^r+Q\sum_j A_jz_j^r.
 \label{eq:defect}
\end{equation}
Define $e_\theta=\theta-\theta^r$, $e_v=v-v^r$, and split mode error as
$z_j-z_j^r=e^{-\lambda_jt}\Delta z_j^0+r_j$, where
$\dot r_j=-\lambda_jr_j+e_v$ and $r_j(0)=0$. The inherited output error is
$b_i(t)=\sum_j a_{ij}e^{-\lambda_jt}\Delta z_{ij}^0$.
Let $U_0(\theta)=\sum_e a_e(1-\cos(B^\top\theta)_e)-p^\top\theta$ and
\begin{align}
 \Phi(\theta,\theta^r)&=U_0(\theta)-U_0(\theta^r)
                         -\nabla U_0(\theta^r)^\top e_\theta,\nonumber\\
 \mathcal V&=\tfrac12e_v^\top Me_v+\Phi
                 +\tfrac12\sum_{i,j}q_ia_{ij}r_{ij}^2.
 \label{eq:movingstorage}
\end{align}

\begin{proposition}[Nonlinear propagation of inherited-force error]
Suppose both edge-angle paths lie in $|\delta_e|<\gamma<\pi/2$ and the
reference momentum defect satisfies $d\in L^2(0,T_f;\R^N)$.
With $m_* =\min_iM_i$ and
$\kappa_\gamma(t)=\tan\gamma\,\|B^\top v^r(t)\|_\infty$, the storage obeys
\begin{equation}
 \dot{\mathcal V}\leq\kappa_\gamma\mathcal V
       +\sqrt{2\mathcal V/m_*}\,\|Qb+d\|.
 \label{eq:movingineq}
\end{equation}
Writing $A_\gamma(t)=\exp[\frac12\int_0^t\kappa_\gamma(s)ds]$, we obtain
\begin{equation}
 \|e_v(t)\|\leq\frac{A_\gamma(t)}{\sqrt{m_*}}
 \left[x_0+\sqrt{T_f/m_*}
          \{\|Qb\|_{L^2}+\|d\|_{L^2}\}\right],
 \label{eq:frequencybound}
\end{equation}
where $x_0=\sqrt{2\mathcal V(0)}$ and both norms are over $[0,T_f]$.
\end{proposition}
\begin{proof}
The kinetic and forced-memory derivatives cancel their cross terms.
The Bregman derivative cancels the nonlinear torque-error term and leaves
\begin{equation*}
 (v^r)^\top[\nabla U_0(\theta)-\nabla U_0(\theta^r)
                      -\nabla^2U_0(\theta^r)e_\theta].
\end{equation*}
The absolute value of each sine remainder is at most
$\frac12\sin\gamma\,|(B^\top e_\theta)_e|^2$, while
$\Phi\geq\frac12\cos\gamma\sum_ea_e|(B^\top e_\theta)_e|^2$.
The remainder is consequently bounded by $\kappa_\gamma\Phi$.
Discarding nonnegative damping gives Eq.~\eqref{eq:movingineq}.
For $x=\sqrt{2\mathcal V}$, a limiting argument at zero gives
$\dot x\leq\kappa_\gamma x/2+\|Qb+d\|/\sqrt{m_*}$.
Variation of constants, $A_\gamma\geq1$, and Cauchy--Schwarz prove the bound.
\end{proof}

No post-event equilibrium is needed for this comparison. The restriction is
a finite-horizon cohesive chart. The next test verifies this condition
using the reference and an uncertainty radius, without assuming that the
unobserved true path is cohesive.
The argument applies to any positive relaxation realization of these
equations; the fractional kernel determines the weights used here.

\subsection{A test for issuing the bound}

Let $L_G=BA B^\top$ and $R_e=B_e^\top L_G^\dagger B_e$ for each active edge.
Effective resistance gives
$|B_e^\top e_\theta|\leq\sqrt{R_e e_\theta^\top L_Ge_\theta}$.
Define the event score
\begin{align}
 S&=\sqrt{e_v(0)^\top Me_v(0)+e_\theta(0)^\top L_Ge_\theta(0)}\nonumber\\
  &\quad+\sqrt{T_f/m_*}\,\|Qb\|_{L^2}.
 \label{eq:eventscore}
\end{align}
The initial square root provides an upper bound on $x_0$ in the chart since
$\nabla^2U_0\preceq L_G$. It is also defined outside that chart.
For a proposed score radius $r$, put
$X_\gamma(t)=A_\gamma(t)[r+\sqrt{T_f/m_*}\,\|d\|_{L^2}]$.
The issuance condition is
\begin{equation}
 |(B^\top\theta^r(t))_e|
          +\sqrt{R_e/\cos\gamma}\,X_\gamma(t)<\gamma
 \label{eq:closure}
\end{equation}
for every active edge and every time in the horizon.
If $S\leq r$, the condition at time zero places the true initial angles inside the
chart, using the score's global phase-error term. At a first exit, the storage
bound, effective resistance, and Eq.~\eqref{eq:closure} keep the
true angle strictly inside, a contradiction. Thus Eq.~\eqref{eq:frequencybound}
holds throughout the horizon whenever this test succeeds and $S\leq r$.

The score is available on a separate labeled calibration ensemble. Suppose the $n_c$ calibration events and the new event are jointly
exchangeable conditional on the frozen training and validation results,
with the same frozen score construction applied to every event. Then
take $k=\lceil(n_c+1)(1-\varepsilon)\rceil$ and the $k$th smallest score
$r_\varepsilon$; use infinity when $k>n_c$.
Split conformal inference gives\cite{lei2018}
\begin{equation}
 \mathbb P\{\text{bound issued and violated at some }t\}\leq\varepsilon.
 \label{eq:jointcoverage}
\end{equation}
Indeed, issuance plus $S\leq r_\varepsilon$ implies validity, and the remaining
event has probability at most $\varepsilon$. This is a marginal joint-failure
statement, not conditional coverage among issued bounds. The predictor and
score must be frozen before calibration. Calibration is separate for each
network/noise group; topology transfer alone provides no coverage guarantee.
The probability is over the calibration sample and a new event, conditional
on the frozen training and validation results.
Several $\gamma$ values can be tested because neither the predictor nor score
changes with $\gamma$.

\subsection{Accounting for numerical propagation}

A solver tolerance alone does not establish an error bound. We construct an
explicit continuous reference from refined forecast velocities: $v^r$ is
linear on each interval, $\theta^r$ is its quadratic integral, and $z^r$
solves the relaxation equation exactly under that linear drive. Only the
momentum equation then has a defect. For constant scheduled injection and
$w=0$ on an interval of length $h$, with midpoint $t_m$ and slope
$s=(v_1^r-v_0^r)/h$, its componentwise upper bound is
\begin{align}
 |d_i(t)|\leq |d_i(t_m)|+\frac h2\left[
 (|B|\diag(a)\omega)_i+D_i|s_i|\right.\nonumber\\
 \left.\qquad+q_i\sum_j a_{ij}
      \max_{\ell=0,1}|v_{i\ell}^r-\lambda_jz_{ij\ell}^r|\right],
 \label{eq:defectsegment}
\end{align}
where $\omega_e=\max_{\ell=0,1}|B_e^\top v_\ell^r|$.
The mode derivative is constant plus one exponential and is monotone on an
interval, so its maximum absolute value occurs at an endpoint.
Squaring the component bounds, summing, and multiplying by interval lengths
bounds $\|d\|_{L^2}^2$. The maximum absolute quadratic angle is bounded by
the larger endpoint angle plus $h^2|B_e^\top s|/8$.
The trapezoidal approximation to the integral of $\|B^\top v^r\|_\infty$
provides an upper bound because
it is a convex function of time on each interval. These quantities test
Eq.~\eqref{eq:closure} over whole intervals, not just sampled times.

For a point forecast $v^p$ different from this reference,
\begin{equation}
 \|v(t)-v^p(t)\|\leq X_\gamma(t)/\sqrt{m_*}
                            +\|v^r(t)-v^p(t)\|.
 \label{eq:transport}
\end{equation}
The known discrepancy transfers a reconstruction-based bound to the
frequency-trained predictor. With nested grids, both velocities are affine
on each refined interval, and their discrepancy norm is at most the larger
endpoint norm. Experiments compare this auxiliary reference with using the
point predictor itself. The analytic guarantees concern the stated finite
equations in real arithmetic. Their evaluation in double precision includes an
arithmetic guard but is not a directed-rounding computer-assisted proof.

\subsection{A reusable critical radius}

The interval enclosures separate the geometry of a reference path from its
calibration radius. Let $\bar\delta_{ke}$ bound its absolute edge angle on
interval $k$, let $\bar A_{\gamma,k+1}$ be the upper growth factor at that
interval's right endpoint, and put $c_d=\sqrt{T_f/m_*}\|d\|_{L^2}$.
The implemented sufficient test is exactly equivalent to
\begin{align}
 r&<r_{\rm crit}(\gamma),\nonumber\\
 r_{\rm crit}(\gamma)&=
 \min_{k,e}\frac{\gamma-\bar\delta_{ke}}
 {\bar A_{\gamma,k+1}\sqrt{R_e/\cos\gamma}}-c_d.
 \label{eq:critical}
\end{align}
Every denominator is positive on a connected graph with positive active
line strengths. Dividing the interval version of Eq.~\eqref{eq:closure}
by that denominator proves the equivalence. Testing a finite set
$\Gamma$ is equivalent to $r<\max_{\gamma\in\Gamma}r_{\rm crit}(\gamma)$.
A nonpositive critical radius rules out any nonnegative calibrated radius
for that reference. This scalar criterion concerns the chosen interval
enclosures, not the necessity of the exact chart condition.

The path and thresholds can be cached before calibration. A new radius
requires comparisons and evaluation of the stored growth factors, without
recomputing the dynamics or angle enclosures. We choose the passing $\gamma$ with
the smallest maximum transported bound; this deterministic choice leaves
the score and its marginal guarantee unchanged.

\paragraph{Training, calibration and deployment.}
(i) From training events, construct the feature scales, linear prior,
integral observer and force decoder. Select regularization and neural
checkpoints using validation alone. (ii) Train the response ridge and its
neural refinement for the point forecast; train a separate force
reconstruction, and freeze both amplitude choices. (iii) Use labeled
calibration events to obtain the order-statistic radius for the frozen
reference. (iv) From a new observation window and the known line event,
infer both memories, propagate the point forecast and refined reference,
and construct the continuous interpolation and defect enclosure.
(v) Calculate Eq.~\eqref{eq:critical}, issue a bound only when an angle threshold passes the test,
and transport the bound using Eq.~\eqref{eq:transport}. Later calibration
sets reuse step (iv). True future trajectories are used in training, validation, calibration,
and retrospective scoring, but are not needed to evaluate the deployment threshold.

%% file: design_table.tex
\begin{table*}[!htbp]
\caption{Essential experimental design. All quantities are nondimensional. Training and validation select the method; the final calibration and test events are newly generated after this selection.}
\label{tab:design}\centering\small
\begin{tabular}{p{.27\textwidth}p{.64\textwidth}}\toprule
Quantity & Design and purpose\\\midrule
Fractional order; modes & $\alpha=0.6$; 24 positive relaxation modes\\
History; observed window; forecast & $[-6,0]$; $[-1,0]$; $[0,2]$\\
Sampling; propagation & $0.05$; bound reference refined to $0.0125$\\
Training; validation & 256; 64 six-node events\\
Final test groups & Six-node 96; ten-cycle 64; ten-mesh 64; twenty-mesh 64\\
Calibration & 64 separate events per network/noise group\\
Observation noise & Standard deviation $0$, $0.002$, $0.01$, $0.03$\\
Decoder dimension & One for clean observations; two for noisy observations, with discarded force energy below $10^{-4}$\\
Calibration miscoverage & $0.1$; order statistic 59\\
Device variation & $M_i\in[0.8,1.2]$, $D_i\in[0.15,0.30]$, $q_i\in[0.25,0.60]$\\
Line strength & $a_e\in[0.8,1.6]$ before a single-line outage that leaves the network connected\\\bottomrule
\end{tabular}\end{table*}

%% file: reproducibility.tex
The operating point is drawn as $\theta_i^*\sim\mathcal N(0,0.12^2)$,
then demeaned, and $p=BA\sin(B^\top\theta^*)$. At history time $-6$,
$\theta_i=\theta_i^*+\mathcal N(0,0.25^2)$,
$v_i\sim\mathcal N(0,0.35^2)$ and every memory mode is zero.
Device and line coefficients are drawn uniformly from the ranges in
Table~\ref{tab:design}. A cycle has the extra chord $(0,N/2)$; meshes add
$N/2$ uniformly chosen distinct missing undirected edges. One uniformly
chosen cycle edge is removed at time zero. Injection is held constant,
and no separate input disturbance is applied. These rules specify the
transient distribution, including its loading and angle margins.
The adaptive backward differentiation formula uses relative and absolute
tolerances $2\times10^{-9}$ and $2\times10^{-11}$ and maximum step $0.12$;
tolerance refinement checks the reference accuracy separately.

At each sample and node the eight features are demeaned phase, frequency,
pre-event and scheduled post-event net injection, $M_i,D_i,q_i$, and time.
Neighbor averaging uses the pre-event weighted adjacency divided by its
row sum. Training means and standard deviations standardize the features;
linear regressions concatenate all local and neighbor time samples.
The neural graph layer lifts the features, concatenates local and neighbor
embeddings, and applies a hyperbolic-tangent map with a residual and layer
normalization. The temporal output is the last token of one causal attention
layer with sinusoidal positions, four heads, feed-forward width twice the
embedding width, rectified-linear activation and zero dropout. A width-48
hyperbolic-tangent head produces the coefficients. Reconstruction and
refinement use embedding width 16 and append noise divided by $0.03$;
refinement also appends the standardized ridge coefficients and initializes
its final residual layer to zero. The direct-state controls use width 32
and output all modes. Their temporal block is selected among the attention
layer, a single gated recurrent unit, and an exponential recurrence with
input-dependent decay. The latter updates $h^+=dh+(1-d)\tanh(Wu)$, with
$d=\exp[-\operatorname{softplus}(a)\Delta t\{1/2+\operatorname{sigmoid}(Gu)\}]$.

All neural models use Adam with learning rate $0.002$, 120 epochs,
batch size 64 and gradient-norm clipping at one. The validation objective
selects the checkpoint. Reconstruction and response models share training
events across the four noise levels and use a known-noise input; the direct
state priors are trained at noise $0.002$. Each level uses one fixed
Gaussian observation perturbation, scaled from the same draws across levels.
The compact models thus use 1024 event--noise examples and 16 batches per
epoch, versus 256 examples and four batches for the direct-state models.
Perturbations are independent across nodes, samples, phase and frequency.
The direct graph-prior row uses unrestricted integral correction. At primary
noise its five selected encoders are three gated recurrent units and two
attention models; at high noise all five use exponential recurrence.
The state-residual control uses width 16 and four covariance coordinates,
trained on standardized residual memory at noise $0.002$. Its validation
criterion weights those residuals by their decaying force; final selection
uses the nonlinear forecast and the resolution rule stated below.
Feature, memory-target and coefficient standard deviations have additive
floors $10^{-5}$, $10^{-4}$ and $10^{-8}$, respectively, to avoid division
by a vanishing scale. All scales are learned from training data.

The linear regularization penalties are selected from
$\{10^{-5},10^{-4},10^{-3},10^{-2},10^{-1},1,10\}$.
Integral corrections compare $L=I$ with ranks $2,4,6,8$, and
$\rho\in\{10^{-7},10^{-6},\ldots,10^2\}$, also allowing no correction.
For noise $(0,0.002,0.01,0.03)$ the selected triples
(prior penalty, correction rank, $\rho$) are
$(10^{-5},I,10^{-7})$, $(10^{-3},4,10^{-2})$,
$(10^{-2},I,10^{-1})$, and $(1,2,10)$.
All response ridges select $10^{-5}$ from the same penalty grid.
The centered response perturbation is $10^{-3}$ in standardized coefficient
units; halving it tests derivative accuracy. Each decoder chooses its final
amplitude from $\{0,1/4,1/2,1\}$ on validation events. The force-coordinate and
principal-direction refinements both select an amplitude of one in every noisy condition. The five force
reconstruction amplitudes are $(1/2,1/4,1/4,1/2,1/4)$ at primary noise and
$(1/2,1/4,1/2,1/2,1/4)$ at high noise. Certification uses the first training
replicate throughout. The state-residual control additionally requires a
validation gain greater than $1.589\times10^{-4}$, its predeclared reference
integration error, and is disabled at primary noise.

Five training initializations are retained, and percentile intervals use 3000 paired
bootstrap resamples of whole events. Bounds test
$\Gamma=\{\pi/6,\pi/4,\pi/3,1.4\}$, a fixed range of chart widths;
selecting a passing width changes neither the score nor calibration rank.
The nonlinear solve uses five Newton updates; equation and energy residuals
check convergence. Linear algebra and propagation use double precision;
neural training and inference use single precision. The reproducibility package accompanying this version contains frozen models,
event seeds, selected settings, training curves and event-level results. It
provides separate procedures for replaying frozen models and retraining in
an empty output directory, with an explicit manifest of the retained
comparator-selection input. The complete package is available under the
Data Availability statement.

%% file: accuracy_table.tex
\begin{table*}[!htbp]
\caption{Mean event frequency root-mean-square error, multiplied by $10^3$. Neural entries average five training repetitions within each event. The paired $95\%$ intervals resample events and condition on these trained models; negative values favor force refinement. All values are nondimensional.}
\label{tab:accuracy}\centering\small
\begin{tabular}{lrrrr}\toprule
Method & Six-node & Ten-cycle & Ten-mesh & Twenty-mesh\\\midrule
\multicolumn{5}{l}{Observation-noise standard deviation $0.002$}\\\midrule
Integral observer, zero prior & 2.991 & 2.920 & 3.203 & 3.117\\
Mean prior, covariance correction & 2.793 & 2.854 & 3.036 & 2.941\\
Linear prior with correction & 2.202 & 2.278 & 2.631 & 2.652\\
Direct graph prior with correction & 2.447 & 2.503 & 2.788 & 2.859\\
State-residual graph prior & 2.202 & 2.278 & 2.631 & 2.652\\
Force-coordinate response ridge & 1.600 & 1.662 & 2.027 & 2.059\\
Principal-direction response ridge & 1.621 & 1.688 & 2.003 & 2.045\\
Principal-direction refinement & 1.570 & 1.619 & 1.929 & 1.976\\
Stand-alone force-coordinate neural predictor & 2.115 & 2.179 & 2.544 & 2.562\\
Force reconstruction objective & 2.193 & 2.272 & 2.621 & 2.644\\
Force-coordinate refinement & 1.506 & 1.558 & 1.952 & 1.985\\
Reference memory, same noisy origin & 2.015 & 1.987 & 2.406 & 2.433\\
\midrule
\multicolumn{5}{l}{Observation-noise standard deviation $0.03$}\\\midrule
Linear prior with correction & 30.319 & 30.924 & 36.512 & 36.779\\
State-residual graph prior & 29.635 & 30.206 & 35.438 & 35.862\\
Force-coordinate response ridge & 18.587 & 18.965 & 21.099 & 21.910\\
Principal-direction response ridge & 18.575 & 18.954 & 21.088 & 21.898\\
Principal-direction refinement & 17.754 & 18.092 & 20.198 & 21.070\\
Stand-alone force-coordinate neural predictor & 21.969 & 22.203 & 25.087 & 26.104\\
Force-coordinate refinement & 17.057 & 17.485 & 19.481 & 20.365\\
\midrule
\multicolumn{5}{l}{Force refinement minus comparator: paired interval, multiplied by $10^3$}\\\midrule
Force ridge, noise $0.002$ & $[-0.131,-0.058]$ & $[-0.138,-0.071]$ & $[-0.115,-0.034]$ & $[-0.106,-0.042]$\\
Principal refinement, noise $0.002$ & $[-0.084,-0.043]$ & $[-0.083,-0.038]$ & $[-0.003,0.049]$ & $[-0.015,0.032]$\\
Force ridge, noise $0.03$ & $[-1.928,-1.162]$ & $[-1.882,-1.109]$ & $[-1.950,-1.284]$ & $[-1.769,-1.314]$\\
Principal refinement, noise $0.03$ & $[-0.900,-0.506]$ & $[-0.819,-0.408]$ & $[-0.892,-0.534]$ & $[-0.823,-0.587]$\\
\bottomrule\end{tabular}\end{table*}

%% file: refinement_results.tex
Figures~\ref{fig:diagnostics}(a,b) show that the differences between the leading methods persist
as the common propagator is refined. At primary noise, the mean improvement
changes by at most $3.93\times10^{-6}$ between the largest and smallest step sizes,
and every interval retains its sign. All five individual training repetitions
improve on the ridge in every group; their finest-step reductions range from
$2.77\%$ to $8.78\%$. 

For the first sixteen six-node test events, the response-surrogate discrepancy
from the nonlinear force forecast is $1.39\times10^{-6}$ at primary noise
and $2.89\times10^{-4}$ at high noise, compared with nonlinear RMSEs of
$1.60\times10^{-3}$ and $1.77\times10^{-2}$. Halving the coefficient
perturbation changes the local response by less than $2.4\times10^{-10}$
in relative norm. The surrogate is local, but its discrepancy is small for
these learned updates, and the final comparisons use nonlinear propagation.

%% file: mechanism_results.tex
\begin{table*}[!tb]
\caption{Six-node inherited-force decomposition and origin-state intervention. Force entries average squared $L^2$ error over events, summed over nodes with $q_i^2$ weights. Frequency RMSE is multiplied by $10^3$ at $h=0.0125$. Exact origins replace only electrical states, retaining memories inferred from noisy windows. Refinements average five training repetitions; reconstruction uses the first.}
\label{tab:mechanism}\centering\small
\begin{tabular}{llrrrrr}\toprule
Method & Noise & Omitted & Coefficient & Total force & Noisy origin & Exact origin\\\midrule
Response ridge & 0.002 & 5.39e-08 & 0.000287 & 0.000288 & 1.597 & 1.669\\
Force refinement & 0.002 & 5.39e-08 & 0.000331 & 0.000331 & 1.506 & 1.797\\
Principal refinement & 0.002 & 3.76e-08 & 0.000384 & 0.000384 & 1.571 & 1.851\\
Force reconstruction & 0.002 & 5.39e-08 & 5.56e-05 & 5.57e-05 & 2.169 & 0.907\\
Response ridge & 0.03 & 5.15e-07 & 0.073 & 0.073 & 18.571 & 24.159\\
Force refinement & 0.03 & 5.15e-07 & 0.0797 & 0.0797 & 17.040 & 24.967\\
Principal refinement & 0.03 & 5.07e-07 & 0.078 & 0.078 & 17.737 & 24.822\\
Force reconstruction & 0.03 & 5.15e-07 & 0.00193 & 0.00193 & 30.244 & 5.526\\\bottomrule\end{tabular}
\end{table*}

Table~\ref{tab:mechanism} resolves the force error into the two terms of
Eq.~\eqref{eq:forcedecomposition}. Coefficient error dominates the omitted
component, so increasing decoder rank would not address the main error here.
Figure~\ref{fig:diagnostics}(c,d) relates that decomposition to prediction:
the refinement reduces frequency error despite a larger force error, and its
advantage is reversed when the origin states are exact. The smaller held-out omitted
component for principal directions also shows that optimality for the training
force map does not prescribe the ordering on new histories.

%% file: certificate_table.tex
\begin{table*}[!tb]
\caption{Ranges across five disjoint calibration sets with a fixed point predictor and test events. Neural uses the auxiliary reconstruction; Direct calibrates the maximum sampled frequency error. Width is the median maximum envelope over issued events; Ratio is the median ratio of maximum envelope to maximum actual vector error. Coverage is the percentage of test scores within the calibration radius. Ranges describe calibration variability, not confidence intervals. Figure~\ref{fig:issuance} compares linear and neural references. Direct radii have frequency units; physical radii use Eq.~\eqref{eq:eventscore}.}
\label{tab:certificate}\centering\small
\begin{tabular}{lllrrrrr}\toprule
Group & Noise & Reference & Issued & Radius & Width & Ratio & Score coverage (\%)\\\midrule
Six-node & 0.002 & Neural & 94--94/96 & 0.025--0.028 & 0.048--0.051 & 9.7--10.2 & 92.7--96.9\\
Six-node & 0.002 & Direct & 96--96/96 & 0.006--0.008 & 0.006--0.008 & 1.2--1.5 & 86.5--97.9\\
Ten-cycle & 0.002 & Neural & 63--63/64 & 0.034--0.038 & 0.065--0.070 & 10.6--11.4 & 89.1--96.9\\
Ten-cycle & 0.002 & Direct & 64--64/64 & 0.008--0.009 & 0.008--0.009 & 1.3--1.4 & 85.9--100.0\\
Ten-mesh & 0.002 & Neural & 64--64/64 & 0.039--0.047 & 0.091--0.102 & 12.3--13.8 & 78.1--95.3\\
Ten-mesh & 0.002 & Direct & 64--64/64 & 0.010--0.011 & 0.010--0.011 & 1.3--1.5 & 87.5--90.6\\
Twenty-mesh & 0.002 & Neural & 63--63/64 & 0.058--0.066 & 0.142--0.153 & 13.5--14.6 & 85.9--96.9\\
Twenty-mesh & 0.002 & Direct & 64--64/64 & 0.013--0.016 & 0.013--0.016 & 1.3--1.6 & 89.1--93.8\\
Six-node & 0.03 & Neural & 76--82/96 & 0.248--0.277 & 0.398--0.440 & 5.8--6.3 & 89.6--95.8\\
Six-node & 0.03 & Direct & 96--96/96 & 0.091--0.106 & 0.091--0.106 & 1.3--1.5 & 88.5--96.9\\
Ten-cycle & 0.03 & Neural & 24--34/64 & 0.310--0.343 & 0.537--0.583 & 5.9--6.2 & 79.7--98.4\\
Ten-cycle & 0.03 & Direct & 64--64/64 & 0.116--0.124 & 0.116--0.124 & 1.3--1.4 & 85.9--95.3\\
Ten-mesh & 0.03 & Neural & 22--30/64 & 0.355--0.382 & 0.661--0.702 & 7.5--7.9 & 87.5--93.8\\
Ten-mesh & 0.03 & Direct & 64--64/64 & 0.118--0.128 & 0.118--0.128 & 1.4--1.5 & 89.1--98.4\\
Twenty-mesh & 0.03 & Neural & 0--0/64 & 0.485--0.518 & --- & --- & 78.1--90.6\\
Twenty-mesh & 0.03 & Direct & 64--64/64 & 0.158--0.170 & 0.158--0.170 & 1.2--1.2 & 90.6--95.3\\\bottomrule\end{tabular}
\end{table*}

%% file: calibration_interpretation.tex
For the high-noise twenty-node reconstruction, the critical radii range from $0.046$ to $0.396$, whereas calibrated radii range from $0.485$ to $0.518$. No test event satisfies the sufficient condition based on these enclosures. Larger phase excursions, accumulated growth and aggregate initial uncertainty consume the available angle margin; this abstention does not demonstrate dynamical instability.

Under the first calibration set, the neural reference yields an issuance rate of $78/96$ in the six-node high-noise group, with a $95\%$ Wilson interval $[72.3,87.8]\%$, and $26/64$ in the ten-cycle group, with an interval of $[29.5,52.9]\%$. These conditional finite-test intervals are distinct from the calibration-set ranges in Table~\ref{tab:certificate}. All group-wise Wilson intervals and event width distributions are included in the complete reproducibility package. Reusing test events across calibration sets does not increase the independent test count. Zero observed physical-envelope failures does not establish conditional coverage among issued events.

%% file: sensitivity_results.tex
\subsection{Changes in memory order, horizon and rate band}

We rebuild the observer and decoder and retrain for each condition in
Table~\ref{tab:sensitivity}, using common event seeds and 128 training,
32 validation, 64 calibration, and 64 test events at noise $0.03$.
This focused study uses one training initialization; its paired intervals
describe event variation conditional on that initialization. The observation
window remains $[-1,0]$. All final choices precede calibration and testing,
and forecasts use step $0.0125$. The wider realization changes both data
generation and prediction, so it tests robustness to a better-resolved kernel,
not transfer of a fixed 24-mode estimator to a different physical system.
\begin{table*}[!tb]
\caption{Sensitivity to fractional order $\alpha$, forecast horizon $T_f$ and number of modes $R$. Frequency errors and the paired force-minus-ridge $95\%$ interval are multiplied by $10^3$. The 72-mode band is $[10^{-7},10^5]$; 24-mode cases use $[10^{-5},10^3]$. Issuance uses the reconstruction reference; $\bar\delta_{\max}$ is the largest reference angle enclosure in radians.}
\label{tab:sensitivity}\centering\small
\begin{tabular}{rrrrrrrrr}\toprule
$\alpha$ & $T_f$ & $R$ & Ridge & Force & Principal & Paired interval & Issued & $\bar\delta_{\max}$\\\midrule
0.6 & 2 & 24 & 19.83 & 18.40 & 18.76 & [-1.94,-0.90] & 49/64 & 1.01\\
0.4 & 2 & 24 & 22.06 & 20.97 & 21.14 & [-1.63,-0.54] & 36/64 & 1.00\\
0.8 & 2 & 24 & 19.86 & 19.17 & 19.18 & [-1.35,-0.02] & 56/64 & 1.03\\
0.6 & 1 & 24 & 21.00 & 20.02 & 20.49 & [-1.62,-0.28] & 63/64 & 0.72\\
0.6 & 3 & 24 & 18.70 & 17.41 & 17.36 & [-1.82,-0.74] & 34/64 & 1.17\\
0.6 & 2 & 72 & 19.55 & 18.23 & 18.62 & [-1.84,-0.78] & 49/64 & 1.00\\\bottomrule\end{tabular}
\end{table*}
Force refinement improves on the response ridge in every tested
condition. Principal refinement is slightly more accurate at the longest
horizon, reinforcing that the force-map optimum is not universal superiority in frequency
prediction. Extending the horizon reduces issuance as both accumulated
uncertainty and the reference angle range grow. The largest angle enclosures
reach approximately one radian, so these comparisons do not reduce to
infinitesimal perturbations about equilibrium. The $\alpha=0.8$ case has
larger omitted high-rate mass: its 24-mode kernel error is $14.2\%$.
Its result concerns that declared realization, not a comparably resolved
continuum model at every order.
\begin{table}[!tb]
\caption{Widening the positive relaxation realization at $\alpha=0.6$. Kernel error is relative $L^1(0,8)$ error in percent. Trajectory RMSE, multiplied by $10^3$, compares 16 shared transients with the 288-mode realization; the latter is a numerical comparison reference.}
\label{tab:kernel}\centering\small
\begin{tabular}{rrrr}\toprule
Modes & Rate band & Kernel error & Frequency RMSE\\\midrule
24 & $[10^{-5},10^{3}]$ & 1.9916 & 4.8753\\
72 & $[10^{-7},10^{5}]$ & 0.3018 & 0.7362\\
144 & $[10^{-9},10^{7}]$ & 0.0469 & 0.1003\\
288 & $[10^{-11},10^{9}]$ & 0.0074 & reference\\\bottomrule\end{tabular}
\end{table}
Table~\ref{tab:kernel} examines the effects of both the omitted rate tails
and interior quadrature. Kernel integration uses a singularity-removing change
of variable with breakpoints at modal time scales, and its absolute integral
is checked against the analytically known signed integral. The simultaneous
decrease in kernel and trajectory errors provides numerical evidence of convergence of this
sequence of finite realizations. The wider 72-mode learning experiment
retains the forecast ordering seen in the base condition. These comparisons
connect the finite calculation to power-law history without claiming an
exact fractional reference or a validated electrical-device model.
The 24-mode discrepancy from the wider reference can exceed the
within-realization forecasting error, so improving the estimator does not
eliminate the need to resolve the hereditary kernel.

%% file: cost_table.tex
\begin{table}[!htbp]
\caption{Per-event deployment time in milliseconds: median [10th, 90th percentiles] over twelve repetitions after one warm-up, in batches of sixteen six-node primary-noise events. One processor thread is used, with single-precision inference and double-precision propagation. Timings describe these implementations.}
\label{tab:cost}\centering\small
\begin{tabular}{lrr}\toprule
Predictor & Initialization & Total\\\midrule
Response ridge & 0.136 [0.135,0.141] & 2.319 [2.308,2.342]\\
Force refinement & 0.296 [0.289,0.309] & 2.485 [2.474,2.513]\\
Force reconstruction & 0.254 [0.243,0.271] & 2.443 [2.423,2.457]\\\bottomrule\end{tabular}
\end{table}

%% file: cached_timing.tex
For eight events and five proposed radii, caching Eq.~\eqref{eq:critical} reduces median evaluation time from $1.225$ to $0.246$ seconds, including path construction in both cases. Five repetitions after one warm-up give a median speed ratio of $5.01$. Forty event--radius comparisons reproduce the issuance decisions and bound values exactly in double precision. The time saving applies to repeated calibration of a fixed path.